\documentclass[useAMS,usenatbib]{mn2e}
\usepackage{amsmath,amssymb}
\usepackage{natbib}
\usepackage{graphicx}
\usepackage{times}
\usepackage{threeparttable}

\def\be{\begin{equation}}
\def\ee{\end{equation}}
\def\bea{\begin{eqnarray}}
\def\eea{\end{eqnarray}}

\title[Fermi GRBs: Interpretation and Implication]{Interpretation and implication of the non-detection of
GeV spectrum excess by Fermi $\gamma$-ray Space Telescope in most
GRBs}
\author[Yi-Zhong Fan]{
Yi-Zhong Fan$^{1,2}$
\thanks{Email: yizhong@nbi.dk}
\\
$^{1}${Niels Bohr International Academy, Niels Bohr Institute,
University of Copenhagen, Blegdamsvej 17, DK-2100
Copenhagen, Denmark} \\
$^{2}${Purple Mountain Observatory, Chinese Academy of Sciences,
Nanjing 210008, China} }

\begin{document}
\date{\today}
\maketitle
\label{firstpage}

\begin{abstract}
Since the launch of the Fermi Gamma-ray Space Telescope on 11 June
2008, significant detections of high energy emission have been
reported only in six Gamma-ray Bursts (GRBs) until now. In this work
we show that the lack of detection of a GeV spectrum excess in
almost all GRBs, though somewhat surprisingly, can be well
understood within the standard internal shock model and several
alternatives like the photosphere-internal shock (gradual
magnetic dissipation) model and the magnetized internal shock model.
The delay of the arrival of the $>100$ MeV photons from some Fermi
bursts can be interpreted too. We then show that with the polarimetry of
prompt emission these models may be
distinguishable. In the magnetized internal shock
model, high linear polarization level should be typical. In the
standard internal shock model, high linear polarization level is
still possible but much less frequent. In the photosphere-internal
shock model, the linear polarization degree is expected to be
roughly anti-correlated with the weight of the photosphere/thermal
component, which may be a unique signature of this kind of model.
We also briefly discuss the implication of the current Fermi GRB
data on the detection prospect of the prompt PeV neutrinos. The
influences of the intrinsic proton spectrum and the enhancement of
the neutrino number at some specific energies, due to the cooling of
pions (muons), are outlined.
\end{abstract}

\section{Introduction}
Gamma-ray Bursts (GRBs) are the most extreme explosion discovered so
far in the universe. With the discovery of the afterglows and then
the measurement of the redshifts in 1997 \citep[see][for a
review]{wijers00}, the cosmological origin of GRBs has been firmly
established. The modeling of the late ($t>10^{4}$ s) afterglow data
favors the external forward shock model \citep[see][for
reviews]{piran99, mesz02, zm04}. The radiation mechanisms employed
in the modeling are synchrotron radiation and synchrotron
self-Compton (SSC) scattering. In the early time the prolonged
activity of the central engine plays an important role in producing
afterglow emission too, particularly in X-ray band
\citep[e.g.,][]{Katz98,FW05,Nousek06,Zhang06}. The radiation
mechanisms, remaining unclear, are assumed to be the same as those
of the prompt soft gamma-ray emission. It is expected that in the
Fermi era the origin of the prompt emission can be better
understood. This is because the Large Area Telescope (LAT) and the
Gamma-ray Burst Monitor (GBM) onboard Fermi satellite
(http://fermi.gsfc.nasa.gov/) can measure the spectrum in a very
wide energy band (from 8 keV to more than 300 GeV), with which some
models may be well distinguished. For example, in the standard
internal shock model the SSC radiation can give rise to a distinct
GeV excess while in the magnetized outflow model no GeV excess is
expected.

Motivated by the detection of some $>100$ MeV photons from quite a
few GRBs by the Compton Gamma Ray Observatory satellite in
1991$-$2000 \citep[e.g.,][]{Hurley94,fm95,Gonz03}, the prompt high
energy emission has been extensively investigated and most
calculations are within the framework of the standard internal
shocks \citep[e.g.,][cf. Giannios
2007]{Pilla98,pw04,gz07,Bosnjak09}. The detection prospect for LAT
seems very promising \citep[see][for a recent review]{fp08}.

Since the launch of Fermi satellite on 11 June 2008, significant
detections of prompt high energy emission from GRBs have been only
reported in GRB 080825C \citep{Bouvier08}, GRB 080916C
\citep{Abdo09}, GRB 081024B \citep{Omodei08}, GRB 090323
\citep{Ohno09b}, GRB 090328 \citep{Cutini09}, possibly and GRB
090217 \citep{Ohno09} until now (5 May 2009). Though the detection
of 3 prompt photons above $10$ GeV from GRB 080916C at redshift
$z\sim 4.5$ \citep{Abdo09,Greiner09} is amazing and may imply a very
high initial Lorentz factor of the outflow $\Gamma_{\rm i}>1800$ and
an efficient acceleration of particles to very high energy
\citep{Zou09}, the non-detection of a significant $>100$ MeV
emission from most GRBs may be a better clue of the underlying
physics. A delay in the onset of the $>100$ MeV emission with
respect to the soft gamma-rays, as detected in GRB 080825C, GRB
080916C, GRB 081024B and GRB 090323, may be the other clue of the
GRB physics \citep{Abdo09,Ohno09b}. In this work we focus on these
two novel observational features.

This work is structured as the following. In section 2 we discuss
the constraint of the current Fermi GRB data on the standard
internal shock model and several alternatives. In section 3 we look
for distinguished signals in linear polarization of the prompt
emission. In section 4 we briefly discuss the implication of the
current Fermi results on the detection prospect of PeV neutrinos
from GRBs. We summarize our results in section 5.


\section{Interpreting the lack of GeV excess in most
GRBs and the delay of the arrival of the $>100$ MeV photons} In the
leading {\it internal shock model} for the prompt emission
\citep{npp92,px94,rm94,dm98}, the ultra-relativistic outflows are
highly variable. The faster shells ejected at late times catch up
with the slower ones ejected earlier and then power energetic
forward/reverse shocks at a radius $R_{\rm int} \sim 5\times
10^{13}~(\Gamma_{\rm i}/300)^2 (\delta t_{\rm v}/{\rm 10~ms})$ cm,
where $\Gamma_{\rm i}$ is the initial Lorentz factor of the outflow
and $\delta t_{\rm v}$ is the intrinsic variability timescale. Part
of the shock energy has been used to accelerate electrons and part
has been given to the magnetic field. If the outflow is magnetized
\citep{usov92,dt92,lb03,gs05}, we call the shocks generated in the
collisions within the outflow the {\it magnetized internal shocks}
\citep{Spruit01,fan04}. The synchrotron radiation of the
shock-accelerated electrons may peak in soft gamma-ray band and then
account for the observed prompt emission. This model has been widely
accepted for the following good reasons: (1) For an
ultra-relativistic outflow moving with an initial Lorentz factor
$\Gamma_{\rm i}$, the velocity (in units of $c$) is $\beta_{\rm
i}=\sqrt{1-1/\Gamma_{\rm i}^2}$. A small velocity dispersion $\delta
\beta_{\rm i}\sim \beta_{\rm i}/(2\Gamma_{\rm i}^2)$ will yield a
very different Lorentz factor. As a result, internal shocks within
the GRB outflow seem inevitable. (2) In the numerical simulation of
the collapsar launching relativistic outflow, people found highly
variable energy deposition in the polar regions in a timescale as
short as $\sim 50$ ms \citep{mw99}. (3) This model can naturally
account for the variability that is well detected in prompt
gamma-ray emission \citep{kps97}. On the other hand, this model
usually predicts a fast cooling spectrum $F_\nu \propto \nu^{-1/2}$
in the X-ray band. However, the data analysis finds a typical X-ray
spectrum $F_\nu \propto \nu^{0}$ \citep{preece00,band93}. Such a
divergence between the model and the observation data is the
so-called ``the low energy spectral index crisis" \citep{gc99}.
Another potential disadvantage of the internal shock model is its
low efficiency of converting the kinetic energy of the outflow into
prompt emission \citep[e.g.,][]{Kumar99}. Among the various
solutions put forward, a plausible scenario is the {\it
photosphere-internal shock model} \citep[e.g.,][]{rm05,peer06,tmr07}.
The idea is that the thermal emission leaking from
the photosphere is the dominant component of the prompt sub-MeV
photons \citep{tho94,mr00,Ryde05,ioka07}. The nonthermal high energy
emission is likely the external inverse Compton (EIC) radiation of
the internal shock-accelerated electrons cooled by the thermal
photons from the photosphere \citep{rm05,peer05,peer06,tmr07}. If
the electrons are accelerated by gradual magnetic energy dissipation
rather than by internal shocks, it is called the {\it
photosphere-gradual magnetic dissipation model} \citep{gian07}.
There is an increasing interest in these two kinds of models since:
(1) In the spectrum analysis people did find evidences for a thermal
emission component in dozens of bright GRBs
\citep{Ryde05,Ryde06,RF09,McGl09}. (2) The emission from the
photosphere can naturally account for the temporal behaviors of the
temperature and flux of these thermal radiation \citep{peer08}. (3)
The overall spectrum of the prompt emission can be reasonably
interpreted \citep[e.g.,][]{peer06,gian07}. (4) The GRB efficiency
can be much higher than that of the internal shock model
\citep[see][and the references therein]{RF09}.

In this section we test these four models with the current Fermi GRB
data. It is somewhat surprisingly to see that none of these models
have been ruled out.

\subsection{Explaining the lack of GeV spectrum excess in most
GRBs}\label{sec:GeV-excess}

\subsubsection{The standard internal shock model} In this
model, the outflow is baryonic and the thermal emission during the
initial acceleration of the outflow is ignorable. The prompt
emission is powered by energetic internal shocks. There are three
basic assumptions. (i) $\epsilon_{\rm e}$, $\epsilon_{\rm B}$,
$\epsilon_{\rm p}$ fractions of shock energy have been given to
electrons,
 magnetic field and protons, respectively (note that
 $\epsilon_{\rm e}+\epsilon_{\rm B}+\epsilon_{\rm
p}=1$). (ii) The energy distribution of the shock-accelerated
electrons is a single power-law. (iii) The prompt soft gamma-ray
emission is attributed to the synchrotron radiation of the shocked
electrons.

For internal shocks generating at $R_{\rm int}$, the typical random
Lorentz factor of the electrons can be estimated as (see section
4.1.1 of \citet{fp08} for details)
\begin{equation}
\gamma'_{\rm e, m} \sim 760~(1+Y_{\rm ssc})^{1/4}L_{\rm
syn,52}^{-1/4} R_{\rm int,13}^{1/2}(1+z)^{1/2}(\varepsilon_{\rm
p}/300~{\rm keV})^{1/2}, \label{eq:gamma_em}
\end{equation}
where $\varepsilon_{\rm p}=h\nu_{\rm m}$ is the observed peak energy
of the synchrotron-radiation spectrum ($\nu F_\nu$), $h$ is the Planck's constant, $L_{\rm syn}$
is the synchrotron-radiation luminosity of the internal shock
emission, $Y_{\rm ssc} \sim {[-1+\sqrt{1+4\epsilon_{\rm
e}/(1+g^{2})\epsilon_{\rm B}}]}/2$ is the regular SSC parameter
\citep{se01,fp08,Tsvi09}, and $g\sim \gamma'_{\rm e,m}
\varepsilon_{\rm p}/\Gamma_{\rm i} m_{\rm e}c^2$. The SSC in the
extreme Klein-Nishina regime ($g\gg 1$) is very inefficient. If that
happens the non-detection of GeV spectrum excess in most Fermi GRBs
can be naturally explained. With the typical parameters adopted
in eq.(\ref{eq:gamma_em}) we have $g\sim 1$, for which the SSC may
still be important (i.e., $Y_{\rm ssc}\geq 1$). In this work the
convenience $Q_{\rm x}=Q/10^{\rm x}$ has been adopted in cgs units
except for some specific notations.

The SSC radiation will peak at
\begin{eqnarray}
h\nu_{\rm m, ssc} &\sim &  {2{\gamma'}_{\rm e, m}^2 \varepsilon_{\rm
p} \over 1+2g} \sim 220 {\rm GeV}~(1+2g)^{-1}(1+Y_{\rm
ssc})^{1/2}\nonumber\\
&& L_{\rm syn,52}^{-1/2} R_{\rm int,13}(1+z)(\varepsilon_{\rm
p}/300~{\rm keV})^{2}. \label{eq:SSC_flare}
\end{eqnarray}
Taking into account the energy
loss of the electrons via inverse Compton scattering on prompt soft
gamma-rays, the cooling Lorentz factor can be roughly estimated as
\begin{equation}
\gamma'_{\rm e,c} \sim 0.03 L_{\rm syn,52}^{-1}R_{\rm int,
13}\Gamma_{2.5}^3.
\end{equation}
In reality $\gamma'_{\rm e,c}$ is always larger than 1.  The derived
$\gamma'_{\rm e,c}<1$ just means that the electrons have lost almost
all their energies and are sub-relativistic.

Prompt high energy photons above the cut-off frequency $\nu_{\rm
cut}$ will produce pairs by interacting with softer photons and will
not escape from the fireball. Following \citet{ls01} and
\citet{fp08}, we have
\begin{eqnarray}
h\nu_{\rm cut} &\approx & 2~{\rm GeV}~(1+z)^{-1}(\varepsilon_{\rm
p}/300~{\rm
keV})^{(2-p)/p}L_{\rm syn,52}^{-2/p}\nonumber\\
&& \delta t_{\rm
v,-2}^{\rm 2/p}\Gamma_{\rm i,2.5}^{(2p+8)/p}. \label{eq:nu_cut}
\end{eqnarray}

The SSC radiation spectra can be approximated by $F_{\nu_{\rm ssc}}
\propto \nu^{-1/2}$ for $\nu_{\rm m}<\nu<\nu_{\rm m,ssc}$, and
$F_{\nu_{\rm ssc}} \propto \nu^{-p/2}$ ($F_{\nu_{\rm ssc}} \propto
\nu^{-p}$) for $\nu>\nu_{\rm m,ssc}$ and $g \leq 1$ ($g\gg 1$). The
energy ratio of the SSC radiation emitted below $\nu_{\rm cut}$ to
the synchrotron radiation in the energy range $\nu_{\rm
m}<\nu<\max\{\nu_{\rm M},~\nu_{\rm cut}\}$ can be estimated as
\begin{eqnarray}
{\cal R} &\sim & {Y_{\rm ssc} \nu_{\rm m}^{(2-p)/2} \int^{\nu_{\rm
cut}}_{\nu_{\rm m}}\nu^{-1/2}d\nu \over \nu_{\rm m,ssc}^{1/2}
\int^{\max\{\nu_{\rm
M},~\nu_{\rm cut}\}}_{\nu_{\rm m}}\nu^{-p/2}d\nu } \nonumber\\
&\approx & (p-2)(\nu_{\rm cut}/\nu_{\rm m,ssc})^{1/2}Y_{\rm ssc},
\end{eqnarray}
where $\nu_{\rm M}\approx 30~\Gamma_{\rm i}(1+z)^{-1}~$ MeV is the
maximal synchrotron radiation frequency of the shocked electrons \citep{cw96}.

\begin{figure}
\includegraphics{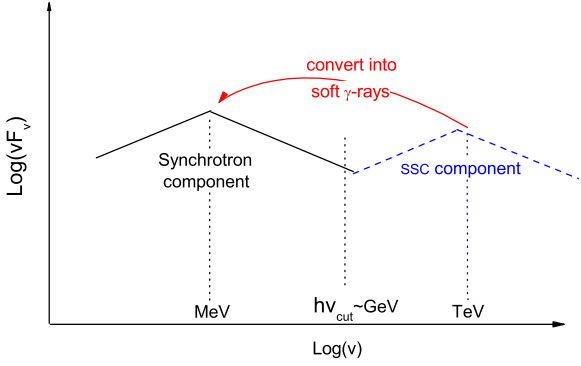}
\caption{A possible interpretation of the non-detection of the SSC
component by Fermi satellite in the internal shock model. The high
energy photons with an energy $>h\nu_{\rm cut}$ have been absorbed
by the soft gamma-rays and energetic $e^{\pm}$ pairs are formed. The
pairs will lost their energy through synchrotron radiation and/or
inverse Compton scattering and then produce soft gamma-rays.}
\label{fig:Schematic-1}
\end{figure}

For $Y_{\rm ssc} \sim 1$, $p\sim 2.5$ (corresponding to the typical
$\gamma-$ray spectrum $F_\nu \propto \nu^{-1.25}$ for
$h\nu>\varepsilon_{\rm p}$) and $\nu_{\rm cut}\ll \nu_{\rm m,ssc}$,
we have ${\cal R}\ll 1$. Therefore there is no GeV excess in the
spectrum, in agreement with the data. In other words, the
non-detection of a significant high energy component is due to a too
large $h\nu_{\rm m,ssc}\sim$ TeV and a relative low $h\nu_{\rm
cut}\sim$ GeV (see Fig.\ref{fig:Schematic-1} for a schematic plot).
The other possibility is that $(1+g^{2})\varepsilon_{\rm
B}>4\epsilon_{\rm e}$, for which $Y_{\rm ssc}\sim {\cal O}(1)$,
i.e., the SSC radiation is unimportant and can be ignored.

\subsubsection{The photosphere$-$internal shock model} \citet{tho94}
proposed the first photosphere model for the prompt gamma-ray
emission, in which the nonthermal X-ray and gamma-ray emission are
attributed to the Compton upscattering of the thermal emission by
the mildly relativistic Alfv\'en turbulence. The spectra of GRBs can
be nicely reproduced \citep[see also][]{peer06,gian07}. It is,
however, difficult to explain the energy dependence of the width of
the gamma-ray pulse \citep{feni95,norr96}. As shown in
\citet{tmr07}, such a puzzle may be solved in the photosphere$-$
internal shocks model, in which the sub-MeV emission is dominated by
the thermal emission of the fireball and the nonthermal tail is the
EIC radiation of the electrons accelerated in internal shocks at a
radius $R_{\rm int}\sim 10^{14}$ cm
\citep[e.g.,][]{mr00,rm05,peer06}. In this model the shock
accelerated-electrons take a power-law energy distribution
$dn/d\gamma'_{\rm e} \propto {\gamma'}_{\rm e}^{-p}$ for
$\gamma'_{\rm e}\geq \gamma'_{\rm e,m} \sim 1$. Such an initial
distribution can not keep if the electrons cool down rapidly. As
shown in \citet{SPN98}, in the presence of steady injection of
electrons, the energy distribution can be approximated by
$N_{\gamma_{\rm e}} \propto {\gamma'}_{\rm e}^{-(p+1)}$ for
$\gamma'_{\rm e}>\max\{\gamma'_{\rm e,c}, ~\gamma'_{\rm e,m}\}$ and
$N_{\gamma_{\rm e}} \propto {\gamma'}_{\rm e}^{-2}$ for
$\gamma'_{\rm e,c}< \gamma'_{\rm e}<\gamma'_{\rm e,m}$. Under what
conditions can shock acceleration generate a particle distribution
with $\gamma'_{\rm e,m} \sim 1$, with a significant fraction of the
outflow energy deposited in the nonthermal particles? There could be
two ways. One is to assume that electron/positron pair creation in
the outflow is so significant that the resulting pairs are much more
than the electrons associated with the protons. As a result, the
fraction of shock energy given to each electron/positron will be
much smaller than that in the case of a pair-free outflow and a
$\gamma'_{\rm e,m} \sim 1$ is achievable \citep{tmr07}. The other
way is to assume that the particle heating is continuous. In the
internal shock scenario, this could happen if an outflow shell
consists of many sub-shells and the weak interaction between these
sub-shells may be able to produce multiple shocks that can
accelerate electrons continually with a very small $\gamma'_{\rm
e,m}$.

Since both $\gamma'_{\rm e,m}$ and $\gamma'_{\rm e,c}$ are $\sim 1$,
the EIC spectrum should be $F_\nu \propto \nu^{-p/2}$ in the MeV-TeV
energy range and there is no GeV excess for $p\sim 2.5$, consistent with the Fermi
data.

\subsubsection{The magnetized internal shock model}\label{sec:mag} If
the unsteady GRB outflow carries a moderate/small fraction of
magnetic field, the collision between the fast and slow parts will
generate strong internal shocks and then produce energetic soft
gamma-ray emission. As usual, the ratio between the
magnetic energy density and the particle energy density is denoted as $\sigma$.
In the ideal MHD limit, for $\sigma\gg 1$ just a very small fraction
of the upstream energy can be converted into the downstream thermal
energy. Therefore the GRB efficiency is very low\footnote{For the
Poynting-flux dominated outflow (i.e., $\sigma\gg 1$),
\citet{usov94} proposed that at the radius $r_{_{\rm MHD}}\sim
6\times10^{15}L_{52}^{1/2} \sigma_{2}^{-1} {t_{v,m}}_{-3}
\Gamma_{\rm i,2.5}^{-1}~{\rm cm}$, the MHD condition breaks down and
large scale electromagnetic waves are generated, where $L$ is the
total luminosity of the outflow and $t_{v,m}$ is the minimum
variability timescale of the central engine \citep[see
also][]{zm02}. The particles are accelerated and the synchrotron
radiation of the ultra-relativistic electrons peaks at $\nu_{\rm
m}\sim 5\times10^{19}\sigma_{2}^{3}(\epsilon_{\rm
e}/0.2)^2[3(p-2)/(p-1)]^{2} \Gamma_{2.5} {t_{v,m}}_{-3}~{\rm Hz}$
~\citep{fzp05}, provided that a significant part of the magnetic
energy has been converted to the thermal energy of the particles. In
this scenario, the SSC radiation is expected to be very weak because
in the rest frame of the electrons having $\gamma'_{\rm e,m} \sim
10^{4}(\epsilon_{\rm e}/0.2)[3(p-2)/(p-1)]\sigma_{2}$
~\citep{fzp05}, the soft gamma-rays have an energy $\gamma'_{\rm
e,m}h\nu_{\rm m}/\Gamma_{\rm i}\gg m_{\rm e}c^2$, i.e., the SSC is
in the extreme Klein-Nishina regime and is very inefficient. The
non-detection of high energy emission from most GRBs is, of course,
consistent with this model.}. That's why people concentrate on the
internal shocks with a magnetization $\sigma\leq 1$
\citep{Spruit01,fan04}.

For the magnetized internal shocks with $\sigma\geq 0.1$, no
significant high energy emission is expected since: (a)
The SSC emission of the internal shocks is weak. Therefore there is
no distinct GeV excess in the spectrum. (b) The
synchrotron-radiation spectrum may be very soft, which renders the
detection of GeV photons from GRBs more difficult. The reason is the
following. For an isotropic diffusion and a relativistic shock, the
electron energy distribution index can be estimated by \citep{KW05}
\begin{equation}
p \sim (3\beta_{\rm u}-2\beta_{\rm u}\beta_{\rm d}^2+\beta_{\rm
d}^3)/(\beta_{\rm u}-\beta_{\rm d})-2. \label{eq:KW05}
\end{equation}
However, in the presence of a large scale coherent magnetic field,
the diffusion is highly anisotropic rather than isotropic
\citep{Morl07a}.  There are thus corrections to eq.(\ref{eq:KW05}).
But as long as the scattering is not very forward- or
backward-peaked, these corrections are small \citep{Keshet06}.
Taking into account the anisotropic correction, \citet{Morl07} found
a spectrum steep to $p\sim 3$ for $\sigma \sim 0.05$. In the
ion-electron shock simulation, the acceleration of particles at the
un-magnetized shock front is a lot more efficient than that with a
$\sigma\sim 0.1$ \citep{Spit06}. Motivated by these two possible
evidences, we adopt eq.(\ref{eq:KW05}) to estimate the spectral
slope of accelerated particles at the magnetized shock fronts. The
validity of our approach can be tested by the advanced numerical
simulations in the future.

\begin{figure}
\begin{picture}(0,180)
\put(0,0){\includegraphics{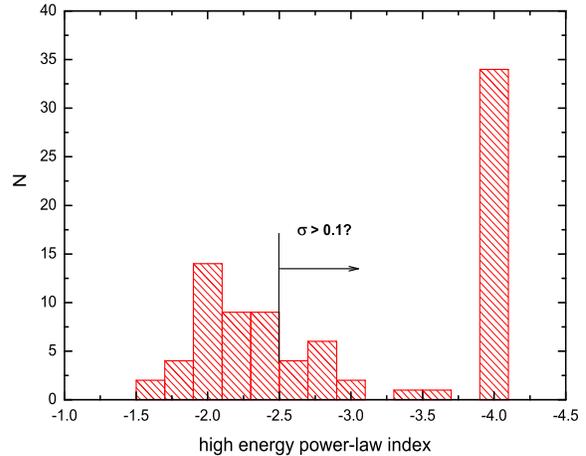}}
\end{picture}
\caption{The distribution of the high energy power-law index
($\beta_{\rm Band}$) of GRBs detected by Fermi satellite from 10
August 2008 to 31 March 2009. The data are taken from
http://www.batse.msfc.nasa.gov/gbm/circulars/ (see also
http://gcn.gsfc.nasa.gov/gcn3$_{-}$archive.html) and are
preliminary. The GRBs having a single power-law spectrum in the GBM
energy range are excluded in the statistics. Following
\citet{preece00} we take $\beta_{\rm Band}=-4$ for all bursts that
can be better fitted by a power-law function with an exponential
high energy cutoff rather than by the Band function \citep{band93}.
We find that a good fraction of bursts have a $p\sim -2(\beta_{\rm
Band}+1)>2.22$, consistent with \citet{preece00}. These soft spectra
may be attributed to the magnetized internal shocks.}
\label{fig:beta}
\end{figure}

In the case of $\sigma=0$, for an ultra-relativistic shock,
$\beta_{\rm u} \rightarrow 1$ and $\beta_{\rm d} \rightarrow 1/3$,
we have $p \rightarrow 2.22$. But for an ultra-relativistic
magnetized shock, $\beta_{\rm u} \rightarrow 1$ and
\citep[e.g.,][]{fan04}
\begin{equation}
\beta_{\rm d} \approx {1\over 6}(1+\chi+\sqrt{1+14\chi+\chi^2}),
\label{eq:fan04}
\end{equation}
where $\chi\equiv \sigma/(1+\sigma)$, please note that $\sigma$ is
measured in the upstream. Note in this work we just discuss the
ideal MHD limit, i.e., there is no magnetic energy dissipation at
the shock front. For $0<\sigma\ll 1$, we have $p\sim
(4.22-2\sigma)/(1-2\sigma)-2>2.22$. For $\sigma\gg1$, we have
$\beta_{\rm d} \rightarrow 1-1/2\sigma$ and $p \sim 4\sigma-1 \gg
2.22$. Correspondingly, the electron spectrum is very soft or even
thermal-like. Adopting $\beta_{\rm u}\sim 1$ and substituting
$\sigma \sim (1,~0.5,~0.1,~0.01)$ into
eqs.(\ref{eq:KW05}-\ref{eq:fan04}), we have $p\sim
(6.6,~4.5,~2.7,~2.3)$. The (very) soft high energy spectra of some
GRBs \citep[e.g.,][see also our Fig.\ref{fig:beta} for the Fermi
GRBs]{preece00} may be interpreted in this way.

If the weak prompt high energy emission of GRBs is indeed attributed
to the magnetization of the outflow, one can expect that the smaller
the $p$, the stronger the high energy emission. The ongoing analysis
of the LAT data will test such a correlation.

\subsubsection{The photosphere$-$gradual magnetic dissipation model}
\citet{gian07} calculated the emission of a Poynting-flux-dominated
GRB outflow with gradual magnetic energy dissipation (reconnection).
In his scenario, the energy of the radiating electrons is determined
by heating and cooling balance. The mildly relativistic electrons
stay thermal throughout the dissipation region because of Coulomb
collisions (Thomson thick part of the flow) and exchange of
synchrotron photons (Thomson thin part). Rather similar to
\citet{tho94}, the resulting spectrum naturally explains the
observed sub-MeV break of the GRB emission and the spectral slopes.
In this scenario, different from the magnetized internal shock
model, the higher the initial $\sigma$, the harder the spectrum (see
the Fig.2 of \citet{gian07} for illustration). For an initial
$\sigma\leq 40$ (corresponding to the baryon loading $L/\dot{M}c^2
\sim \sigma^{3/2} \leq 250$, where $\dot{M}$ is the mass loading
rate), the resulting $>10$ MeV spectrum is very soft \citep[see
also][]{Drenkhahn02,DS02}, accounting for the failed detection of
the GeV spectrum excess in most GRBs.

\subsection{Interpreting the delay of the arrival of
the $>100$ MeV photons}\label{sec:GeV-delay}

In both the collapsar and the compact star merger models for GRBs
\citep[see][for reviews]{piran99,mesz02,zm04}, the early outflow may
suffer more serious baryon pollution and thus have a smaller
$\Gamma_{\rm i}$ than the late ejecta \citep[][]{ZhangW04}. This may
explain the delay of the arrival of the $>100$ MeV emission since as long as
\[
\Gamma_{\rm i}\leq \Gamma_{\rm i,c}=180~(1+z)^{p \over
2p+8}({h\nu_{\rm cut} \over 100~{\rm MeV}})^{p\over
2p+8}({\varepsilon_{\rm p}\over 300~{\rm keV}})^{p-2 \over
2p+8}L_{\rm syn,52}^{1 \over 4+p}\delta t_{\rm v,-2}^{-{1\over
4+p}},
\]
{\it the $>100$ MeV photons can not escape from the emitting region
freely and thus can not be detected.}

In the photosphere-gradual magnetic dissipation model, a small
$\Gamma_{\rm i}$ implies a low initial magnetization of the outflow,
for which the high energy spectrum can be very soft
\citep{DS02,gian07}. In the magnetized internal shock model, the
delay of the onset of the LAT observation indicates a larger magnetization of the early
internal shocks if $\Gamma_{\rm i}>\Gamma_{\rm i,c}$.

In the collapsar scenario, before the breakout, the initial outflow
is choked by the envelope material of the massive star (Zhang et al.
2004). The ultra-relativistic reverse shock may be able to
smooth out the velocity/energy-density dispersion of the initial
ejecta. So the internal shocks generated within the early/breakout
outflow may be too weak to produce a significant non-thermal
radiation component. The early emission is then dominated by the
thermal component from the photosphere and may last a few seconds
(provided that the chocked material has a width comparable to that
of the envelope of the progenitor). The outflow launched after the
breakout of the early ejecta can escape from the progenitor freely
and the consequent internal shocks can be strong enough to produce
energetic non-thermal radiation. The photosphere-internal shock
model therefore might be able to naturally account for the delay in
the onset of the LAT observation.

In summary, before and after the onset of the $>100$ MeV emission,
it seems the physical properties of the outflow have changed.


\section{The linear polarization signal of the prompt $\gamma-$ray emission}
As discussed in section 2, the failed detection of the GeV spectrum
excess in most GRBs can be understood in either the standard
internal shock model or several alternatives. Therefore we need
independent probes to distinguish between these scenarios. Our
current purpose is to see whether the polarimetry in gamma-ray band
can achieve such a goal. In this section, we firstly investigate the
linear polarization property of the photosphere-internal shock model
({\it the results may apply to the photosphere-gradual magnetic
dissipation model as well}) since it has not been reported by others
yet\footnote{In the final stage of preparing the manuscript, the
author was informed by K. Toma and X. F. Wu that the POET
(Polarimeters for Energetic Transients) group discussed the
polarization property of the photosphere-internal shock model
independently. The preliminary results were included in their white
paper.}. We then briefly discuss the linear polarization signals
expected in the magnetized internal shock model and in the standard
internal shock model since they have been extensively discussed in
the literature
\citep[e.g.,][]{Lyu03,Granot03,Waxman03,NPW03,FXW08,Toma09}.

\subsection{Linear polarization signal of the photosphere-internal shock model}
\label{sec:Lin-thermal}
\begin{figure}
\includegraphics{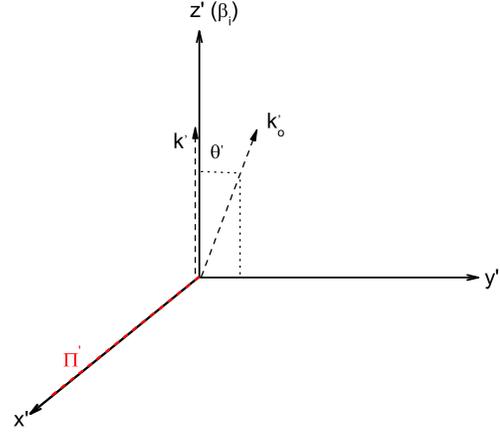}
\caption{The coordinates in the comoving frame of the emitting
region. The polarization vector $\hat{\Pi}'$ is along the ${\rm
X'}-$direction for a positive $P$ given in eq.(\ref{eq:AA81-b})
otherwise it would be along $\hat{\rm X}'\times \hat{k}'_{\rm o}$. }
\label{fig:coordinate-2}
\end{figure}
In this work we assume an uniform outflow.  At any point in the
outflow there is a preferred direction, the radial direction, in
which the fluid moves. We choose the ${\rm Z}'$-direction of the
fluid local frame coordinate to be in that direction. The ${\rm
Y}'-$direction is chosen to be within the place containing the line
of sight (i.e., the scattered photon $k'_{\rm o}$) and the ${\rm
Z}'-$axis (see Fig.\ref{fig:coordinate-2}). In this frame, the
incident photons ($k'$) are along the ${\rm Z}'-$direction.

As usual, we assume that the electrons are isotropic in the comoving
frame of the emitting region\footnote{In the current scenario, the
electrons are heated by the internal shocks and an isotropic
distribution in the rest frame of the emitting region may be a good
approximation. Such a distribution, however, will be modified by the
cooling of the electrons via EIC scattering on the thermal photons
from the inner part of the outflow. Detailed numerical simulation is
needed to see such a modification and then its influence on the
inverse Compton radiation, which is beyond the scope of this work.}.
The incident photons are the thermal emission from the photosphere
and are unpolarized. The energy of the incident and the scattered
photons are $h\nu'_{\rm se}$ and $h\nu'$, respectively. As shown in
\citet{AA81} and \citet{Bere82}: (I) In the cloud of the isotropic
electrons of energies $\gamma'_{\rm e}m_{\rm e}c^2$, the spectrum of
photons, upscattered at the angle $\theta'$ relative to the
direction of the seed photon beam, can be approximated by
\begin{eqnarray}
{d N_\gamma \over  dt d\nu'd\Omega'}& \approx &{3\sigma_T c \over
16\pi {\gamma'}_e^2}{n_{\nu'_{\rm se}} d\nu'_{\rm se} \over
\nu'_{\rm se}} {x\over 2 \beta'_{\rm e}Q} (4A_0^2-4A_0+B_0),
\label{eq:AA81-a}
\end{eqnarray}
where $\beta'_{\rm e}=(1-1/{\gamma'}_{\rm e}^2)^{-1/2}$ and
\begin{eqnarray}
A_0 &=& {1\over 2\delta {\gamma'}_{\rm e}^2 x}({1\over {\cal
T}_1}-{1\over {\cal
T}_2}), \nonumber\\
B_0 &=& {{\cal T}_2\over {\cal T}_1}+{{\cal T}_1\over {\cal
T}_2},\nonumber\\
{\cal T}_1 &=& (1-\cos \theta'){1+x+x\delta \cos \theta'-\delta x^2 \over Q^2},\nonumber\\
{\cal T}_2 &=& (1-\cos \theta'){1+x-x\delta \cos \theta'+\delta \over Q^2},\nonumber\\
Q &\equiv & \sqrt{1+x^{2}-2x\cos \theta'},\nonumber\\
\delta &\equiv & h \nu'_{\rm se}/({\gamma'}_{\rm e} m_{\rm e}
c^{2}),~~x\equiv \nu'/\nu'_{\rm se}.
\end{eqnarray}
Please note that $x$ ranges from $x_{\rm m}$ to $x_{_{\rm M}}$ that
are given by
\begin{eqnarray}
x_{\rm m,_{\rm M}}=1+{\{{\cal A} \mp {\gamma'}_{\rm e}{\beta'}_{\rm
e}\sqrt{{\gamma'}_{\rm e}^2(1+\delta)^{2}(1-\cos
\theta')^{2}+\sin^{2}\theta'}\}\over 1+2\delta {\gamma'}_{\rm e}^2
(1-\cos \theta')+{\gamma'}_{\rm e}^2\delta^2(1-\cos \theta')^{2}},
\end{eqnarray}
where ${\cal A}\equiv (1-\cos \theta'){\gamma'}_{\rm
e}^2[{\beta'}_{\rm e}^2-\delta -\delta^2 (1-\cos \theta')]$.

(II) The polarization degree is
\begin{equation}
P \approx {4(A_0-A_0^2) \over B_0-4A_0+4A_0^2}. \label{eq:AA81-b}
\end{equation}

For the relativistic electrons (i.e., $\beta_{\rm e}\rightarrow 1$),
eq.(\ref{eq:AA81-a}) and eq.(\ref{eq:AA81-b}) take the simplified
forms
\begin{eqnarray}
{d N_\gamma \over dt d\nu' d\Omega'} &\approx & {3\sigma_T c \over
16\pi {\gamma'}_e^2}{n_{\nu'_{\rm se}} d\nu'_{\rm se} \over
\nu'_{\rm se}} [1+ {\xi^2 \over 2(1-{\xi})}-{2\xi \over b_\theta
(1-\xi)}\nonumber\\
&+&{2\xi^2 \over b_\theta^2 (1-\xi)^2}], \label{eq:AA81-a1}
\end{eqnarray}
\begin{equation}
P\approx {{2\xi \over b_\theta (1-\xi)}-{2\xi^2 \over b_\theta^2
(1-\xi)^2}\over 1+ {\xi^2 \over 2(1-{\xi})}-{2\xi \over b_\theta
(1-\xi)}+{2\xi^2 \over b_\theta^2 (1-\xi)^2}}, \label{eq:AA81-b1}
\end{equation}
where $\xi \equiv h\nu'/({\gamma'}_{\rm e} m_{\rm e} c^2)$,
$b_\theta=2(1-\cos \theta'){\gamma'}_{\rm e} h\nu'_{\rm se}/(m_{\rm
e} c^2)$, and $h\nu'_{\rm se}\ll h\nu' \leq {\gamma'}_{\rm e} m_{\rm
e} c^2 b_\theta /(1+b_\theta)$.

Since the emitting region is moving relativistically, the angle
$\theta'$ (in Fig.\ref{fig:coordinate-2}) corresponding to the line
of sight (L.o.S) is given by
\begin{equation}
\cos \theta'=(\cos \theta-\beta_{\rm i})/(1-\beta_{\rm i} \cos
\theta),
\end{equation}
where $\theta$ is the angle between the line of sight and the
emitting point (measured in the observer's frame).

\begin{figure}
\includegraphics{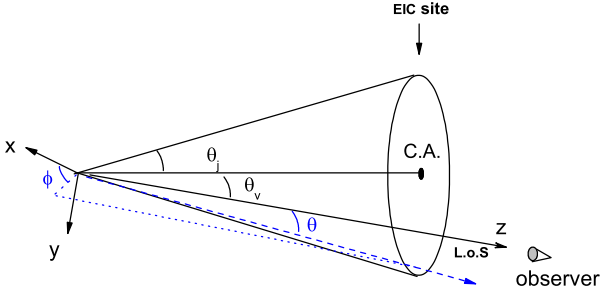}
\caption{Sketch of the geometrical set-up used to compute the
polarization signal. We take the L.o.S as the $z-$axis. The ${\rm
y}$-axis ($x-$axis) is within (perpendicular to) the plane
containing the line of sight and central axis of the ejecta.}
\label{fig:Cartoon}
\end{figure}

The azimuthal angle $\phi$ varying from $0$ to $2\pi$ is defined in
Fig.\ref{fig:Cartoon}. The polar angle $\theta$ ranges from $0$ to
$\theta_{\rm v}+\theta_{\rm j}$. The angle between the vector $(\sin
\theta \cos \phi,~\sin \theta \sin \phi,~\cos \theta)$ and the
central axis of the ejecta (C.A. in Fig.\ref{fig:Cartoon}) is
denoted as $\Theta$ and is given by
\begin{equation}
\cos \Theta =-\sin \theta_{\rm v}\sin \theta \sin \phi+\cos
\theta_{\rm v}\cos \theta.
\end{equation}
Please {\it bear in mind} that in the following radiation
calculation, the flux is set to be zero if $\cos \Theta<\cos
\theta_{\rm j}$ because these points $(\theta,\phi)$ are outside of
the cone of the ejecta.

The EIC radiation flux in the observer frame is
\begin{equation}
F_{\nu,\rm EIC}\propto
\int{ {\cal D}^3 h\nu'{d
N_\gamma \over  dt d\nu'd\Omega'} N_{\gamma_{\rm e}}d{\gamma'}_{\rm
e} d\Omega},
\end{equation}
where $\nu={\cal D}\nu'/(1+z)$, ${\cal D}=[\Gamma_{\rm
i}(1-\beta_{\rm i} \cos \theta)]^{-1}$ is the Doppler factor,
and $\Omega$ is the solid angle satisfying $d\Omega=\sin \theta d\theta
d\phi$.

The polarized radiation flux is
\begin{equation}
Q_{\nu,\rm EIC}\propto
\int{ {\cal D}^3 h\nu' P
\cos 2 \phi {d N_\gamma \over  dt d\nu'd\Omega'} N_{\gamma_{\rm
e}}d{\gamma'}_{\rm e} d\Omega}.
\end{equation}
The polarization degree of the EIC emission is
\begin{equation}
P_{\rm \nu, EIC}=|Q_{\rm \nu, EIC}|/F_{\nu,\rm EIC}.
\end{equation}

\begin{figure}
\begin{picture}(0,200)
\put(0,0){\includegraphics{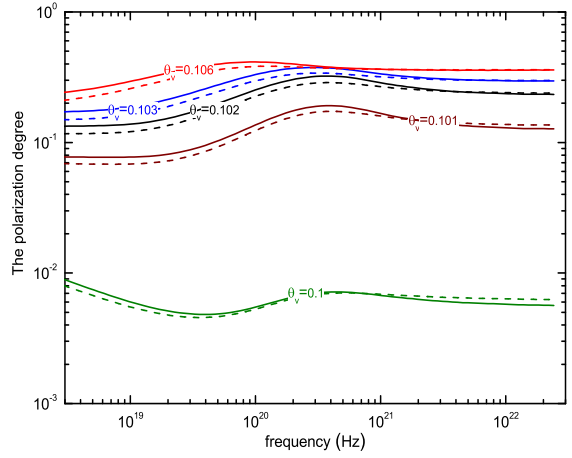}}
\end{picture}
\caption{The degree of the linear polarization of the EIC component
as a function of frequency. The half-opening angle of the ejecta
$\theta_{\rm j}$ and the bulk Lorentz factor $\Gamma_{\rm i}$ are
assumed to be $0.1$ and $300$, respectively. The viewing angle
$\theta_{\rm v}$ is marked in the plot. The solid and dashed lines
are for the electron distribution $N_{\gamma_{\rm e}}\propto
{\gamma'_{\rm e}}^{-3.5}$ and $N_{\gamma_{\rm e}}\propto
{\gamma'_{\rm e}}^{-2}$, respectively.} \label{fig:Pol-1}
\end{figure}

One can see that a non-zero net polarization is expected as long as
$\theta_{\rm v}>0$. In the numerical example, we assume that the
seed photons have a thermal spectrum (as suggested in the
photosphere model)
\begin{equation}
n_{\nu'_{\rm se}}\propto {(h\nu'_{\rm se})^{2} \over e^{h\nu'_{\rm
se}/kT'}-1},
\end{equation}
where $kT'\approx kT/2\Gamma_{\rm i}$ is the temperature (measured
in the rest frame of the emitting region) of the thermal emission.
In the calculation we take $\Gamma_{\rm i} \sim 300$ and $kT \sim
100$ keV. The electron distribution is taken as $N_{\gamma_{\rm e}}
\propto {\gamma'}_{\rm e}^{-(1+p)} \propto {\gamma'}_{\rm e}^{-3.5}$
for ${\gamma'}_{\rm e}>2$, otherwise $N_{\gamma_{\rm e}} =0$. For
comparison purpose we also consider the case of $N_{\gamma_{\rm e}}
\propto {\gamma'}_{\rm e}^{-2}$.  The numerical results are
presented in Fig.\ref{fig:Pol-1}. One can see that the polarization
degrees expected in these two representative cases are only slightly
different. We also find that a moderate linear polarization level
($P_{\rm \nu,EIC}>10\%$) is achievable only for $\theta_{\rm
v}\gtrsim \theta_{\rm j}+1/(3\Gamma_{\rm i})$.

\begin{figure}
\begin{picture}(0,200)
\put(0,0){\includegraphics{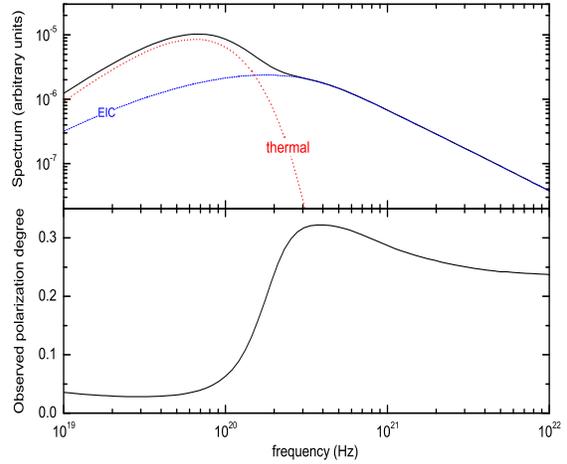}}
\end{picture}
\caption{The spectrum of one simulated GRB (the upper panel) and the
expected linear polarization degree of the total emission as a
function of the photon energy (the lower panel). In the calculation
we take $\theta_{\rm v}=0.102$. Other parameters are the same as
those used in Fig.\ref{fig:Pol-2} in the case of $N_{\gamma_{\rm
e}}\propto {\gamma'_{\rm e}}^{-3.5}$.} \label{fig:Pol-2}
\end{figure}

Currently the prompt emission consists of a thermal and a
non-thermal components. The thermal component with the flux $F_{\rm
\nu,th}$ is expected to be unpolarized while the nonthermal EIC
component may have a high linear polarization level. The observed
polarization degree
\[
P_{\rm \nu,obs}={|Q_{\rm \nu,EIC}|\over F_{\rm \nu,EIC}+F_{\rm
\nu,th}}
\]
should be strongly frequency-dependent. Roughly speaking, the linear
polarization degree is anti-correlated with the weight of the
thermal component. With an energy $\sim kT$, the emission is
dominated by the thermal component and $P_{\rm kT,obs}$ is low. For
$h\nu \gg kT$, the emission is dominated by the EIC component and
$P_{\nu,\rm obs} \sim P_{\nu,\rm EIC}$, as illustrated in
Fig.\ref{fig:Pol-2}. This unique behavior\footnote{For the
synchrotron radiation of the electrons moving in a random magnetic
field (the standard internal shock model) or in an ordered magnetic
filed (e.g., the magnetized internal shock model), before and after
the peak of the spectrum, the polarization degree changes because
the polarization properties depend on the profile of the spectrum.
However, such a dependence is weak, as shown in \citet{Granot03}.}
can help us to distinguish it from other models.

The probability of detecting a
moderate/high linear polarization degree (${\cal R}_{\rm pol}$), however, is not high (Please note that
we do not take into account the weak events for which a reliable polarimetry is impossible). On
the one hand, a high linear polarization level is achievable only for
$\theta_{\rm v} \geq \theta_{\rm j}+1/3\Gamma_{\rm i}$. On the other
hand, $\theta_{\rm v}-\theta_{\rm j}\lesssim 1/\Gamma_{\rm i}$ is
needed otherwise the burst will be too weak to perform the gamma-ray
polarimetry. For $\Gamma_{\rm i}\theta_{\rm j}\gg 1$, we have
\begin{equation}
{\cal R}_{\rm pol} \sim 4 /(3\Gamma_{\rm i} \theta_{\rm j}) \approx
5\% ~\Gamma_{\rm i,2.5}^{-1}\theta_{\rm j,-1}^{-1}. \label{eq:R_pol}
\end{equation}

During the revision of this work, \citet{McGl09} reported their
analysis on the spectrum and the polarization properties of GRB
061122. They found out that the spectrum was better fitted by the
superposition of a thermal and a non-thermal components and the
photons in the ``thermal" emission dominated energy range had a
(much) lower polarization level than those in the higher energy
band. These two characters are in agreement with the
photosphere-internal shock model (or the photosphere-gradual magnetic dissipation model).

\subsection{Linear polarization level expected in the standard internal
shock model} In the standard internal shock model, the polarization
of the synchrotron radiation depends on both the {\it poorly known}
configuration of magnetic field generated in internal shocks and the
geometry of the visible emitting region. Assuming a random magnetic
field that remains planar in the plane of the shock,
\citet{Waxman03} and \citet{NPW03} showed that a high linear
polarization level can be obtained when a narrow jet is observed
from the edge, like in the photosphere-internal shock model. For the
jets on-axis ($\theta_{\rm v}\leq \theta_{\rm j}$), the linear
polarization degree is low (see also Gruzinov 1999 and Toma et al.
2009). The detection probability of a moderate/high linear
polarization degree can also be estimated by eq.(\ref{eq:R_pol}).

\subsection{High linear polarization degree expected in the magnetized internal shock model}
For the magnetized internal shock model, the prompt soft
$\gamma-$ray emission is attributed to the synchrotron radiation of
the electrons in ordered magnetic field and a high linear
polarization level is expected \citep{Lyu03,Granot03}. The physical
reason is the following. The magnetic fields from the central engine
are likely frozen in the expanding shells. The toroidal magnetic
field component decreases as $R^{-1}$, while the poloidal magnetic
field component decreases as $R^{-2}$. At the radius of the
``internal'' energy dissipation (or the reverse shock emission), the
frozen-in field is dominated by the toroidal component. For an
ultra-relativistic outflow, due to the relativistic beaming effect,
only the radiation from a very narrow cone (with the half-opening
angle $\leq 1/\Gamma_{\rm i}$) around the line of sight can be
detected. As long as the line of sight is off the symmetric axis of
the toroidal magnetic field, the orientation of the viewed magnetic
field is nearly the same within the field of view. The synchrotron
emission from such an ordered magnetic field therefore has a
preferred polarization orientation (i.e. perpendicular to the
direction of the toroidal field and the line of sight).
Consequently, the linear polarization of the synchrotron emission of
each electrons could not be effectively averaged out and the net
emission should be highly polarized. The detection prospect of
a high linear polarization degree is very promising
(i.e., ${\cal R}_{\rm pol}\sim 100\%$). The above argument applies
to the reverse shock emission as well if the outflow is magnetized
\citep{fan04a}.\\

\begin{table}
\caption{Distinguish between the models with the polarimetry data.}
\begin{tabular}{l|c|c|c|c|c}
\hline model & unique polarization property & ${\cal R}_{\rm pol}$
\\ \hline
standard internal shocks &  & $\lesssim$ 10\%
\\ \hline
photosphere-internal shocks$^\dagger$ & strongly frequency-dependent$^\ddagger$
& $\lesssim$ 10\%
\\ \hline
magnetized internal shocks &  & $\sim$ 100\%
\\ \hline
\\
\multicolumn{3}{p{0.45\textwidth}}{\footnotesize $^\dagger${In the photosphere-gradual
magnetic dissipation model, very similar polarization properties are expected.}}\\
\multicolumn{3}{p{0.45\textwidth}}{\footnotesize $^\ddagger${As shown
in Fig.\ref{fig:Pol-2}, the polarization degree is anti-correlated
with the weight of the photosphere/thermal component.}}
\end{tabular}
\label{tab:stat}
\end{table}

As summarized in Tab.\ref{tab:stat}, for the magnetized
internal shock model, a high linear polarization level should be
typical, while for two other models a moderate/high linear
polarization degree is still possible but much less frequent. So the
statistical analysis of the GRB polarimetry results may be able to
distinguish the magnetized internal shock model from the others
\citep[see aslo][]{Toma09}. In the photosphere-internal shock model
the polarization degree is expected to be strongly
frequency-dependent. Such a remarkable behavior, if detected, labels
its physical origin.

Indeed there were some claims of the detection of high linear
polarization degree in the soft $\gamma-$ray emission of GRB 021206
\citep[][however see Rutledge \& Fox 2004]{CB03}, GRB 930131, GRB
960924 \citep{Willis05}, GRB 041219A \citep{McGl07,Gotz09}, and GRB
061122 \citep{McGl09}. These results are consistent with each other
as the errors are very large. The situation is inconclusive and
additional data is needed to test these results. Measuring
polarization is of growing interest in high energy astronomy. New
technologies are being invented, and several polarimeter projects
are proposed, such as, in the gamma-ray band, there are the Advanced
Compton Telescope Mission \citep{Boggs06}, POET \citep{Hill08} and
others \citep[see][for a summary]{Toma09}. So in the next decade
reliable polarimetry of GRBs in gamma-ray band may be realized and
we can impose tight constraint on the models. At present, the most
reliable polarimetry is in UV/optical band
\citep[e.g.,][]{Covi99,Wijers99}. The optical polarimetry of the
prompt emission and the reverse shock emission require a quick
response of the telescope to the GRB alert. This is very
challenging. Mundell et al. (2007) reported the optical polarization
of the afterglow, at 203 sec after the initial burst of
$\gamma-$rays from GRB 060418, using a ring polarimeter on the
robotic Liverpool Telescope. Their robust ($90\%$ confidence level)
upper limit on the percentage of polarization, less than $8\%$,
coincides with the fireball deceleration time at the onset of the
afterglow. Such a null detection is, however, not a surprise because
for this particular burst the reverse shock emission is too weak to
outshine the unpolarized forward shock emission \citep{JF07}. Quite
recently, the robotic Liverpool Telescope performed the polarimetry
measurement of the reverse shock emission of GRB 090102 (Kobayashi 2009, private communication).
Following \citet{fan02}, \citet{zhang03} and \citet{KP03}, it is
straightforward to show that the reverse shock of GRB 090102 is
magnetized. Consequently the optical flash is expected to be highly
polarized. If confirmed in the ongoing data analysis, the magnetized
outflow model for some GRBs will be favored.


\section{Implication on the detection prospect of PeV neutrino emission}
The site of the prompt $\gamma-$ray emission may be an ideal place
accelerating protons to ultra-high energy \citep{Vietri95,Waxman95}.
These energetic protons can produce high-energy neutrinos via
photomeson interaction, mainly through $\Delta$-resonance
\citep{WB97}. The resulting neutrinos have a typical energy $E_{\rm
\nu,obs} \sim 5\times 10^{14}~{\rm eV}~\Gamma_{2.5}^2
[(1+z)^2\epsilon_{\gamma, \rm obs}/1~{\rm MeV}]^{-1}$. Significant
detections are expected if GRBs are the main source of ultra-high
energy cosmic rays \citep{WB97}. The underlying assumption is that
the proton spectrum is not significantly softer than $dN/dE\propto
E^{-2}$. The current Fermi observations do not provide an
observational evidence for such a flat particle spectrum. Below we
discuss the detection prospect of PeV neutrinos implicated by the
non-detection of high energy emission from most GRBs. In the
magnetized outflow model, the acceleration of a significant part of
protons to energies $\geq 10^{16}$ eV is highly questionable because
of the resulting soft proton spectrum. In the photosphere-internal
shock model, $\gamma'_{\rm e,m}\sim 1$ is needed \citep{tmr07}. The
efficiency of accelerating protons to very high energy depends on
the mechanism of the particle heating. For example, in the case of
multiple internal shocks, each pair of internal shocks are expected
to be very weak since $\Gamma_{\rm sh}-1 \sim 0.04 (\gamma'_{\rm
e,m}/5)(\epsilon_{\rm e}/0.2)^{-1}[3(p-2)/(p-1)]^{-1}$, where
$\Gamma_{\rm sh}$ is the Lorentz factor representing the strength of
the shock ($\Gamma_{\rm sh}\sim 1$ for Newtonian shocks). So the
acceleration of the protons to ultra-high energy is less efficient
than the standard internal shocks. This is particularly the case if
the acceleration is mainly via second-order Fermi process, in which
the acceleration of particles depends on the shock velocity
sensitively.

Even in the standard internal shock model, the generation of
$10^{20}$ eV protons and the production of PeV-EeV neutrinos may be
not as promising as that claimed in most literature adopting a
proton spectrum $dN/dE \propto E^{-2}$. Such a flat
spectrum is predicted for {\it Newtonian} shocks and has been
confirmed by the supernova remnant observations. But the typical MeV
spectrum $F_\nu \propto \nu^{-1.25}$ \citep{preece00} of GRBs
suggests $dN/dE\propto E^{-2.5}$, supposing the accelerated protons
and electrons have the same spectrum. The non-detection of $>100$
MeV photon emission from most GRBs implies soft electron (possibly)
and proton spectra. Given a proton spectrum $dN/dE\propto E^{-2.22}$
that is predicted in the relativistic shock acceleration model (the
First-order Fermi mechanism), the kinetic energy of the ejecta needs
to be $\sim 100$ times of the $\gamma-$ray radiation energy if GRBs
are indeed the main source of the observed $\sim 10^{20}$ eV cosmic
rays \citep[see][and the reference therein]{Dermer08}. In other
words, the GRB efficiency should be as low as $\sim 1\%$. If
correct, the number of protons at $E\sim 10^{16}$ eV will be quite a
few times what assumed in \citet{Guetta04}. Correspondingly the PeV
neutrino flux will be higher. However, the current afterglow
modeling usually yields a typical GRB efficiency $\sim 10\%$
\citep[e.g.,][]{FP06,Zhang07} or larger
\citep[e.g.,][]{PK01,Granot06}. Below we discuss a new
possibility--The proton spectrum is curved. In the ``low-energy"
part, the spectrum may be steepened significantly by the leakage of
the very high energy cosmic rays from the ejecta \citep[see][and the
references therein]{Hillas05}. The ``high-energy" spectrum part may
be a lot flatter. For example, in the numerical simulation of cosmic
rays accelerated in some supernova remnants, a spectrum $dN/dE
\propto E^{-1.7}$  at the high energy part is obtained
\citep[e.g.,][]{Volk02,Bere03}. If holding for GRBs as well and GRBs
are the main source of the $10^{20}$ eV cosmic rays, the PeV
neutrino spectrum will be harder than that predicted in
\citet{Guetta04}. For instance, the neutron spectra
$\varepsilon_{\rm \nu}^2dN/d\varepsilon_{\nu} \propto
\varepsilon_{\nu}^0$ and $\varepsilon_{\rm
\nu}^2dN/d\varepsilon_{\nu} \propto \varepsilon_{\nu}^{-2}$ in their
Fig.3 will be hardened by a factor of $\varepsilon_{\nu}^{0.3}$. But
the total flux may be just $\sim 10\%$ times that predicted in
\citet{Guetta04} because in this scenario the protons are not as
many as that suggested in a flat spectrum $dN/dE\propto E^{-2}$ for
$E\ll 10^{20}$ eV.

\begin{figure}
\includegraphics{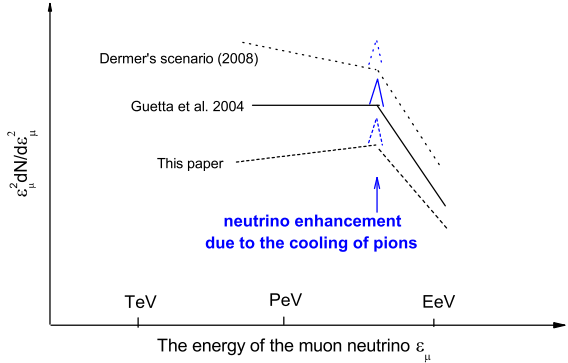}
\caption{The schematic plot of the PeV muon neutrino spectrum in the
standard internal shock model.} \label{fig:Schematic}
\end{figure}

There is a process, ignored in some previous works, that can enhance the detection
prospect a little bit. After the pions (muons) are generated, the
high energy pions (muons) will lose energy via synchrotron radiation
before decaying, thus reducing the energy of the decay neutrinos
(e.g. Guetta et al. 2004). As a result, above
$\varepsilon_{\nu_{\mu}}^{\rm c} \sim{10^{17} \over
1+z}\epsilon_{\rm e}^{1/2}\epsilon_{\rm
B}^{-1/2}L_{\gamma,52}^{-1/2}\Gamma_{\rm i, 2.5}^{4}\delta t_{\rm
v,-2}~{\rm eV}$ ($\varepsilon_{\bar{\nu}_{\mu},\nu_{\rm e}}^{\rm
c} \sim \varepsilon_{\nu_{\mu}}^{\rm c}/10$), the slope of the
corresponding neutrino spectrum steepens by 2, where $L_{\gamma}$ is
the luminosity of the $\gamma-$ray emission \citep{Guetta04}.
However we do not suggest a smooth spectral transition around
$\varepsilon_{\nu_{\mu}}^{\rm c}$ or
$\varepsilon_{\bar{\nu}_{\mu},\nu_{\rm e}}^{\rm c}$ because the
cooling of pions (muons) will cause a pile of particles at these
energies (see also Murase \& Nagataki 2006). A simple estimate suggests that the number of neutrinos in
the energy range  $(0.5,~1)\varepsilon_{\nu_{\mu}}^{\rm c}$ should
be enhanced by a factor of $\sim 3$. A schematic plot of the muon
neutrino spectrum in the standard internal shock model is shown in
Fig.\ref{fig:Schematic}.

\section{Conclusion}
In the pre-Fermi era, it is widely expected that significant
GeV emission will be detected in a good fraction of bright GRBs if
they are powered by un-magnetized internal shocks
\citep[e.g.,][]{Pilla98,pw04,gz07,fp08}. The detection of a distinct
excess at GeV-TeV energies, the SSC radiation component of such
shocks, will be a crucial evidence for the standard fireball model.
The non-detection of the GeV spectrum excess in almost all Fermi
bursts \citep{Abdo09} is a surprise but does not impose a tight
constraint on the models. For example, in the standard internal
shock model, the non-detection can be attributed to a too large
$h\nu_{\rm m,ssc} \sim $ TeV and a relative low $h\nu_{\rm cut}
\sim$ GeV. Some alternatives, such as the photosphere-internal shock
model, the magnetized internal shock model and the
photosphere-gradual magnetic dissipation model, can be in agreement
with the data, too (see Tab.\ref{tab:sum} for a summary).
We attribute the delay in the onset of LAT detection in quite a few
Fermi bursts to the unfavorable condition for GeV emission of the
early outflow (see section \ref{sec:GeV-delay} for details).

\begin{table}
\caption{The physical reasons for the lack of GeV spectrum excess in most GRBs.}
\begin{tabular}{l|c|c|c|c}
\hline model & the physical reason
\\ \hline
standard internal shocks &  $h(\nu_{\rm m,ssc},~\nu_{\rm cut})\sim (\rm TeV,~GeV)$
\\ \hline
photosphere-internal shocks & very small $\gamma'_{\rm e,m}$ and $\gamma'_{\rm e,c}$
\\ \hline
magnetized internal shocks &  soft synchrotron spectrum and weak SSC
\\ \hline
photosphere-gradual \\
magnetic dissipation & very small $\gamma'_{\rm e,m}$ and $\gamma'_{\rm e,c}$
\\ \hline
\end{tabular}
\label{tab:sum}
\end{table}

With the polarimetry of GRBs people can potentially distinguish between
some prompt emission models (see Tab.\ref{tab:stat} for a summary;
see also Toma et al. 2009). We show in section \ref{sec:Lin-thermal}
that in the photosphere-internal shock model the linear polarization
degree is roughly anti-correlated with the weight of the thermal
component and will be highly frequency-dependent. Such a unique
behavior, if detected, labels its physical origin. However, a
moderate/high linear polarization level is expected only when the
line of sight is outside of the cone of the ejecta (i.e.,
$\theta_{\rm v}>\theta_{\rm j}$). In addition, $\theta_{\rm
v}-\theta_{\rm j}\lesssim 1/\Gamma_{\rm i}$ is needed otherwise the
burst will be too weak to perform the gamma-ray polarimetry.
Consequently the detection prospect is not very promising.

In this work we have also briefly discussed the detection prospect
of prompt PeV neutrinos from GRBs. The roles of the intrinsic
spectrum of the protons and the cooling of pions (muons) have been
outlined. The latter always increases the neutrino numbers at the
energies $\varepsilon_{\nu_{\mu}}^{\rm c}$ or
$\varepsilon_{\bar{\nu}_{\mu},\nu_{\rm e}}^{\rm c}$ by a factor of
3. The former, however, is uncertain. If the protons have an
intrinsic spectrum $dN/dE\propto E^{-2.22}$ and have a total energy
about tens times that emitted in gamma-rays, the detection prospect
would be as good as, or even better than that presented in
\citet{Guetta04}. If the proton spectrum traces that of the
electrons, i.e., typically $dN/dE\propto E^{-2.5}$, the detection
prospect would be discouraging.

\section*{Acknowledgments}
We thank the anonymous referee for very helpful suggestions/comments
and Drs. X. F. Wu, K. Toma, and Y. C. Zou for communication. This
work was supported in part by the Danish National Science
Foundation, Chinese Academy of Sciences, National basic research
program of China (grant 2009CB824800), and the National Natural
Science Foundation of China (grant 10673034).

\end{document}